\documentstyle[bibnorm,psfig]{lamuphys}

\makeatletter

\let\chapter\hid@chapter

\makeatother

\begin{document}

\psfull

\title {The Muonium Atom as a Probe of Physics beyond the Standard Model}

\author{ L.~Willmann$^*$ and K.~Jungmann}
\authorrunning{ L.~Willmann and K.~Jungmann}
\institute{ Physikalisches Institut der Universit\"at 
Heidelberg,\\ Philosophenweg 12,  D-69120 Heidelberg, Germany 
}

\maketitle

\abstract{
The observed interactions between particles are not fully explained in the
successful theoretical description of the standard model to date. 
Due to the close confinement of the bound state muonium ($M = \mu^+ e^-$)  
can be used as an ideal probe 
of quantum electrodynamics and weak interaction and also for a search for
additional interactions between leptons.
Of special interest is the lepton number violating process of sponteanous
conversion of muonium to 
antimuonium.
%



\section{Introduction}

Precision measurements on atomic systems have played an important role in the
course of the
development of modern physics. In many cases they have lead to discoveries
which had significant impact on the understanding of the physical laws of nature. 
The explanation of the carefully measured electromagnetic
spectrum of atomic hydrogen by the Schr\"odinger equation 
was a great success for quantum mechanics.  
The observed fine structure was included in the solutions to the Dirac equation
which demonstrated the necessity of a
relativistic description of the atomic structure.
Precise investigations of the hydrogen
Balmer-$\alpha$ line revealed a faint nearby line \cite{urey32} 
which was the discovery of deuterium through its spectroscopic isotope shift. 
A small deviation of the measured hyperfine splitting in hydrogen \cite{rabi} 
from the value predicted in Fermi's theory on the 0.1\,\% level could be
explained by the anomalous magnetic moment of the electron. This discovery
together with the observation of the Lamb-shift
($2^2S_{\frac{1}{2}}-2^2P_{\frac{1}{2}}$) in
 hydrogen \cite{lamb} has initiated and 
boosted the development of the modern field theory of
quantum electrodynamics (QED).

The unification of the weak and electromagnetic interactions 
in the electroweak standard model was strongly supported by the observation   
of parity violation in precise spectroscopic measurements in heavy atoms. 
Today electroweak processes examined both at
high energies \cite{marciano} at the LEP electron-positron storage ring collider
at CERN, Geneva, 
Switzerland, and
in atomic parity violation experiments
in heavy atoms, e.g. in cesium and thallium 
\cite{sandars,wieman}, have ascertained the power of
the unified electroweak theory which is valid over a range 
of 10 orders of magnitude in momentum transfer.
Today the standard model 
appears to be  a very successful effective description of all known 
interactions between particles and no significant deviation from it could be 
established so far.
Its predictions 
are subject to high precision experiments which allow to extract a set
of intrinsic
parameters including the  masses in the leptonic and the quark sectors and
mixing angles between different quarks.
However, there still remain unresolved
questions within this sophisticated theoretical framework 
like the number of interactions, 
the number of lepton and quark generations or
the nature of parity violation. 
Particularly the question of lepton number conservation is investigated by various 
experiments,
since no underlying symmetry could be discovered to be associated with it yet.

The muonium atom (M=$\mu^+ e^-$),
the bound state of a positive muon $\mu^+$ and an electron $e^-$,
can be considered a light hydrogen isotope.
This fundamental system is ideally suited for investigating bound  state quantum 
electromagnetic theory and it renders the possibility to
test fundamental concepts of the standard model.
The spectroscopic measurements of electromagnetic transitions
like the hyperfine interval in the ground
state or the 1s-2s energy splitting are generally considered precise 
tests of QED and are used to infer accurate values of fundamental 
constants \cite{hugheszp}. In addition, they may be used to extract
information on fundamental symmetries. For example, 
the latest measurement of the 1s-2s energy interval \cite{hugheszp,m1s2s}
can be regarded as 
the best test of the charge equality of leptons from different particle 
generations at a 
level of $10^{-8}$ relative accuracy \cite{chrgequ}. 
The system  offers further unique possibilities to search
for yet unknown interactions between leptons, in particular, since 
the close confinement of the bound state 
allows its constituents a rather long interaction time 
which is ultimately limited 
by the lifetime $\tau_\mu = 2.2 \mu s$ of the muon. 

\section{Test of Lepton Number Conservation }

A spontaneous conversion of muonium into antimuonium (${\rm
\overline{M}} = \mu^- e^+)$~ would violate additive lepton family (generation)
number conservation by two units. This process is not provided in the
standard model like others which are intensively searched for, e.g.
$\mu \rightarrow e\gamma$ \cite{mega}, $\mu \rightarrow
eee$ \cite{bert85}, $ \mu -e$ conversion \cite{hone96} or the muon decay mode 
$\mu^+ \rightarrow  e^+ + \nu_\mu + \overline{\nu_e}$ \cite{free93}.
However, in the framework of many speculative theories, which try to 
extend the standard model in order to 
explain some of its not well understood features,
lepton number violation is a natural process and
muonium to antimuonium conversion is an essential part in several 
of these models (Fig. 1) [14-19].      
 
Traditionally, muonium-antimuonium
conversion is described as an effective four fermion interaction 
with a coupling constant ${ G_{\rm M\overline{M}}}$
which can be measured in units of the Fermi coupling constant 
of the weak interaction ${ G_{\rm F}}$ \cite{fein61}.
Many of the speculative models would allow a strength
of the interaction as large as
the experimental bound at the time they were created.

In minimal left-right symmetric theory 
muonium and antimuonium could be coupled through a 
doubly charged Higgs boson $\Delta^{++}$. In this case even a
lower bound has been predicted for  ${ G_{\rm M\overline{M}}}$,
provided the muon neutrino mass
$m_{\nu_\mu}$ were larger than 35 keV/c$^2$ \cite{herc92}.
With the present experimental limit of
$m_{\nu _\mu} \leq $ 170 keV/c$^2$ \cite{assa94}
the coupling constant ${ G_{\rm M\overline{M}}} $
should be larger than 
$2\cdot 10^{-4} ~G_{\rm F}$. This figure would even increase for 
an improved bound on $m_{\nu_\mu}$.

For neutrinos being  Majorana particles a coupling between
muonium and antimuonium is possible by an
intermediate pair of neutrinos. 
A limit on the effective coupling can be estimated based
on the Majorana mass limit of the electron neutrino which has been 
deduced from experimental
searches for neutrinoless double $\beta$-decay \cite{klapdor} 
to ${ G_{\rm M\overline{M}}} \le
10^{-5}{ G_{\rm F}}$ \cite{halp82}.

Supersymmetric theories allow an interaction to be
mediated by a $\tau$-sneutrino
$\tilde\nu_\tau$, the supersymmetric partner of the
$\tau$-neutrino. The predictions are
${ G_{\rm M\overline{M}}} \leq 10^{-2} G_{\rm F}$ for a 
mass value $m_{\tilde\nu_{\tau}}$ of
$100~$GeV/c$^2$ \cite{halp93}.

Some of these speculative theories need to introduce
neutral scalar bosons to 
explain the mass spectrum in the leptonic sector.
These models predict a coupling strength of $10^{-2} G_{\rm F}$ which is in the
range of the sensitivity of present experimental search \cite{wong94}.

In the framework of grand unification theories (GUT) 
muonium-anti\-mu\-onium conversion could be
explained by the exchange of a gauge boson $X^{++}$ 
which carries both an electronic and a
muonic lepton number. Bhabha scattering experiments at PETRA storage 
ring at DESY in Hamburg, Germany, bounded the mass of
this dileptonic particle to $m_{X^{++}}/g_{3l} \ge 340$ GeV/c$^2$, where $g_{3l}$ 
depends on the particular symmetry and is of order unity \cite{petra}. This can be
translated into $G_{\rm M\overline{M}} \le 10^{-2} G_{\rm F}$.

\begin{figure}[t]
\label{fgraph}
\unitlength 1.0 cm 
 \begin{picture}(11.8,4.0) 
  \centering{ 
   \hspace{0.3 cm} 
   \raisebox{-0.6 cm}{
   \mbox{ 
    \psfig{figure=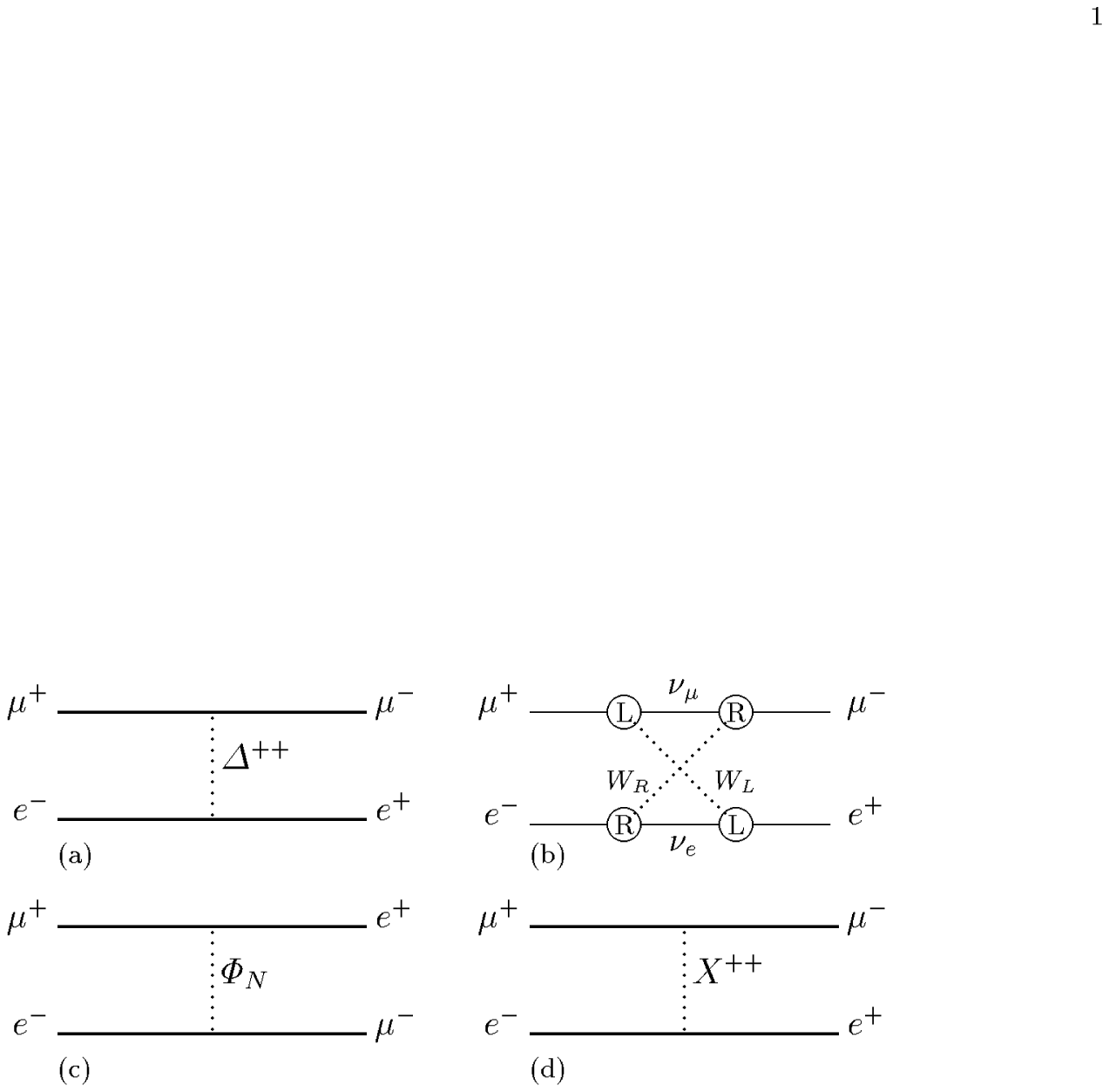,height=11.5cm} 
        }
        }
             } 
 \end{picture} 
  \centering\caption[]
        {
        Muonium-antimuonium conversion can be described in various
        theories beyond the standard model. The interaction
        could be mediated by
        (a) a doubly charged Higgs boson $\Delta^{++}$ \protect{\cite{herc92}},
        (b) heavy Majorana neutrinos \protect{\cite{halp82}},
        (c) a neutral scalar $\Phi_N$ \protect{\cite{wong94}}, which could be 
                for example a supersymmetric $\tau$-sneutrino $\tilde{\nu}_{\tau}$
                \protect{\cite{halp93}} or
        (d) a dileptonic gauge boson $X^{++}$ \protect{\cite{fuji94}}.}
\end{figure}

\section{The Conversion Process}

Muonium and antimuonium are neutral atoms which
are degenerate in their energy levels in the absence of external
fields. In 1957 Pontecorvo suggested the possibility of a muonium to 
antimuonium  conversion process even before the atom had been formed for 
the first time  by V.W. Hughes and his cowor\-kers in 1960 at the NEVIS 
cyclotron of Columbia University, New York, USA \cite{hugh60}.
He proposed a coupling by an intermediate neutrino pair state in analogy 
to the  $ K^\circ -\overline{K^\circ} $ oscillations, which were discovered 
at that time \cite{pontecorvo}.

Any possible coupling between muonium and its antiatom  will give rise to
oscillations between the two species. For atomic s-states with 
principal quantum number $n$
a splitting of their
energy levels
\begin{eqnarray} 
 \delta =
 {\displaystyle \frac{8 G_{\rm F}}{\sqrt{2} n^2 \pi a_0^3}
 \frac{G_{\rm M\overline{M}}}{G_{\rm F}} } 
\end{eqnarray}
is caused,
where $a_0$ is the Bohr radius of the atom.
For the ground state $ \delta $ equals   
$ 1.5 \cdot 10^{-12} \, {\rm eV} \cdot (G_{\rm M\overline{M}}/G_{\rm F})$
which corresponds to 519 Hz for $G_{\rm M\overline{M}}=G_{\rm F}$.
We note that this value is three times larger than the uncertainty
reported for the best measurement of the muonium ground state 
hyperfine structure interval \cite{mari82}.
Therefore, the interpretation of precise measurements of the hyperfine structure
must include considerations on how such a process would affect the accuracy 
under the particular experimental conditions \cite{jung95}.

A system starting at time $t = 0$ as a pure state of muonium could be observed
in the antimuonium state at a later time $t$ with a probability of
(Fig. \ref{oscill})
\begin{eqnarray}
\label{t_oscill}
{ p_{\rm M\overline{M}}(t)} &= \sin^2\left(\frac{\delta \, t}{2\,\hbar}\right)
e^{-\lambda _\mu t} 
&\approx \left(\frac{\delta \, t}{2\,\hbar}\right)^2 \cdot e^{-\lambda _\mu t},
\end{eqnarray}
where $ \lambda _\mu = 1/\tau_\mu $ is the muon decay rate.
The approximation is valid for a weak coupling as suggested by the 
known experimental
limits on $G_{\rm M\overline{M}}$.
In this case the process should be considered a conversion rather than an oscillation.
The maximum of the probability for a decay
as antimuonium is found  at $t_{\rm max}  
= 2 /\lambda_\mu$, while the ratio of antimuonium to
muonium continuosly increases with time.
The total conversion probability integrated over all decay times is 
\begin{eqnarray}
\label{pmm_to_gmm}
{P_{\rm M\overline{M}} = {2.56\cdot10^{-5}}}\left(
\frac{G_{\rm M\overline{M}}}{G_{\rm F}}\right)^2.
\end{eqnarray}
This demonstrates the advantage of experiments 
in which the system is allowed an extended time interval 
(a duration of order $\tau_{\mu}$ or longer)  for developing a conversion.

\begin{figure}[bt]
\label{oscill}
\unitlength 1.0 cm 
 \begin{picture}(11.8,5.2) 
  \centering{ 
   \raisebox{-0.5cm}{
   \mbox{ 
    \psfig{figure=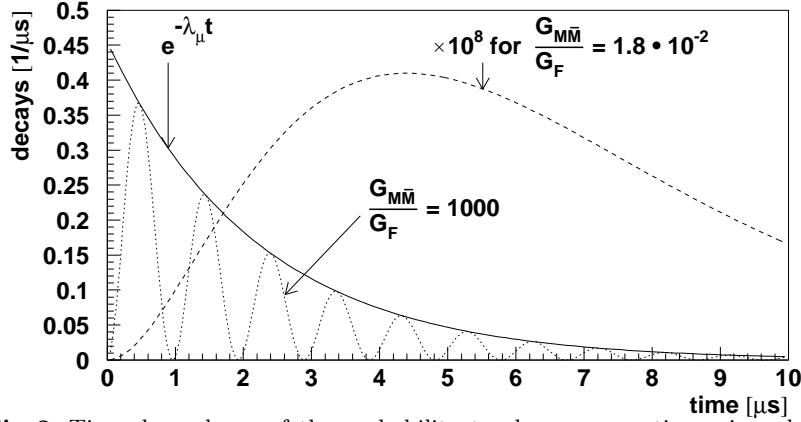,height=5.5cm} 
        }
             } 
                    }
 \end{picture} 
 \caption{
Time dependence of the probability to observe an antimuonium decay 
for a system which was initially in a pure muonium state. 
The solid line represents the exponential decay of muonium 
in the absence of a finite coupling.
The decay probability as antimuonium is given for a coupling strength of
G$_{\rm M\overline{M}}$ = 1000 by the dotted line and for a coupling strength
small compared to the muons decay rate (dashed line).
In the latter case the maximum of the probability
is at 2 muon lifetimes. 
Only for strong coupling 
several oscillation periods could be observed.
}
\end{figure}

The degeneracy of corresponding states in the atom and its antiatom
is removed by external magnetic and
electric fields which can cause a suppression 
of the conversion and a reduction of the probability $p_{\rm M\overline{M}}$.
The influence of an external magnetic field depends on the interaction type of
the process. The reduction of the conversion
probability has been calculated for all possible
interaction types as a function of field strength 
(Fig. \ref{mbar_in_fields}) \cite{wong95,hori95}. 
In the case of an observation of the conversion process  
the coupling type could be revealed by  measurements of
the conversion probability at two different 
magnetic field values.

The conversion process is strongly suppressed for muonium in contact with matter,
since a transfer of the negative 
muon in antimuonium to any other atom is energetically
favored and breaks
up the symmetry between muonium and antimuonium by opening up 
an additional decay channel for the antiatom only.
In gases at atmospheric pressures the conversion probability
is about
five orders of magnitude smaller than in vacuum \cite{morg70}
mainly  due to scattering of the atoms from gas molecules. 
In solids the reduction amounts to even 10 orders of magnitude. 
Therefore a sensitive experiment will benefit largely from employing 
muonium atoms in vacuum. \\

\begin{figure}[tb]
\label{mbar_in_fields}
 \unitlength 1.0 cm 
 \begin{picture}(11.8,5.2)
 \centering { 
   \hspace{0.5 cm} 
   \raisebox{-0.6 cm}{
   \mbox{ 
    \psfig{figure=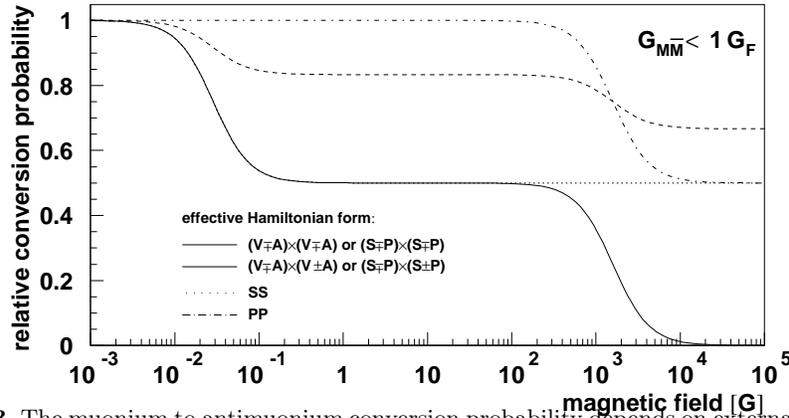,height=5.5cm} 
        }
    }
 } 
 \end{picture}
 \caption{The muonium to antimuonium conversion probability
          depends on external magnetic fields
          and the coupling type. Recent independent
          calculations were performed  by Wong and Hou
          \protect{\cite{wong94}} 
        and Horikawa and
          Sasaki \protect{\cite{hori95}}.
          }
\end{figure}

\section{History of Experimental Search}

There is a strong connection between 
experimental searches for muonium to antimuonium conversion
and the development of
efficient sources of muonium atoms.
In the earliest experiment in 1967 at the NEVIS cyclotron \cite{amat68}
muons were stopped
in a Ar noble gas target of pressure 1 atm with 
a technique similar to the one which was used in the discovery of muonium.
A large fraction of the muons forms muonium by electron capture.
A conversion process would be indicated by 
K$_\alpha$ X-rays originating from an argon atom
after the transfer of the negative muon from antimuonium.
A sensitivity of
$G_{\rm M\overline{M}}$  $< 5800\,G_{\rm F} $ (95\% C.L.) could be reached
which was mainly limited
by the strong suppression of the conversion in gases.

One year later  M{\o}ller scattering was investigated at
the Princeton-Stan\-ford electron
storage rings at Stanford, USA. An analysis  
of the channel  $e^- + e^- \rightarrow \mu ^- + \mu ^- $, which is
essentially the same physical process as muonium-antimuonium conversion,
yielded nearly two orders of magnitude higher sensitivity on the coupling strength 
\cite{barb69}.
\footnote{Today $e^- + e^-$  scattering experiments 
would have to run for approximately 1 year at LEP
beam energies and luminosities to reach a sensitivity similar to a modern
muonium-antimuonium conversion experiment in medium energy laboratories.}

Significant increases in sensitivity could be achieved after 1980 
by taking advantage of newly developed sources of muonium in vacuum.
At the Los Alamos Meson Physics Facility (LAMPF) in Los Alamos, USA, 
muonium in vacuum could be produced by
a beam foil technique from thin aluminum foils.
The velocity-resonant nature of the 
electron capture process causes typical 
kinetic energies of the atoms of a few keV.
The corresponding high velocities and finite dimensions of the apparatus
restricted the time interval available for the conversion.
Antimuonium could have been discovered by secondary electrons and muonic X-rays
from a bismuth catcher foil. The coupling constant 
$G_{\rm M\overline{M}}$ could be limited to below  $ 7.5\,G_{\rm F}$ (90\% C.L.)
\cite{ni93}.

The discovery of muonium formation  
in fine grain SiO$_2$ powders \cite{mars78,cab88} 
with about 60 \% efficiency \cite{kief82},
after stopping muons from a surface beam stimulated
experimental work at the Tri University Meson 
Facility (TRIUMF) in Vancover, Canada, in the early 80's.
The signature for antimuonium was the detection of X-rays after
capturing the negative muon in a calcium host atom.
The data were analyzed under the assumption that the muonium atoms escape
from the grains into the intergranular voids
and yielded $G_{\rm M\overline{M}}$~$\leq$~42\,$G_{\rm F}$ 
(95\% C.L.) \cite{mars82}. 
With a more complete understanding of the behaviour of muonium atoms
inside of a powder target
a reanalysis of the data limited the coupling 
constant $G_{\rm M\overline{M}}$ to less than 20\,$G_{\rm F}$
(95\% C.L) \cite{beer86}.

The observation of a few percent of the muonium atoms  
leaving SiO$_2$ powder target surfaces
with thermal energies at TRIUMF \cite{beer86} and at
at the Paul Scherrer Institut (PSI) in Villigen, Switzerland
\cite{wood88},
was a major breakthrough in the mid 80's.
It has boosted experimental efforts searching for muonium to 
antimuonium conversion and has been employed in all 
new approaches since.
 
At TRIUMF an experiment using thermal
muonium in vacuum requested a signature consisting
of X-rays generated by the transfer of the negative muon to a host atom
and followed by the delayed decay of a radioactive tantal nucleus created
by nuclear muon capture.  
A sensitivity of  $G_{\rm M\overline{M}} \le 0.29\,G_{\rm F}$ (90\% C.L.)
could be reached \cite{hube90}.

The thermal kinetic energy of the muonium atoms 
corresponds to a velocity of 7.4(1) mm/$\mu$s \cite{wood88}. This
assures that the atoms will stay in a small volume 
of about 100 cm$^3$ for several natural lifetimes $\tau_{\mu}$
and allows for long times for the conversion to antimuonium. 

At the Phasotron accelerator in Dubna, Russia, another experiment has been carried out
using muonium in vacuum from a SiO$_2$ target. 
The only signal required was the  observation of 
a single  energetic electron from the negative muon's decay in a narrow momentum band
of $6.3$ MeV/c right below the maximum possible momentum 53 MeV/c of the 
decay electron in muon decay.  
A limit of $G_{\rm M\overline{M}} \le 0.14\,G_{\rm F}$ (90\% C.L.) 
was deduced after one single event has been observed to fulfill the weak required
criterion \cite{gord94}. 

\section{Coincidence Signatures of the Atom's Decay}

Major progress was achieved using a new powerful and clean
signature requesting the coincident
detection of both constituents of the antimuonium atom in its decay.
This method was developed and applied for the first time in an approach 
at LAMPF, where a magnetic spectrometer was used to
search for an energetic electron from the $\mu^-$ decay.
The positron, which is expected to be left behind 
from the atomic shell with a mean kinetic energy corresponding to the system's
Rydberg energy \cite{chat92},
could be electrostatically extracted from the interaction region
and registered on a microchannel plate (MCP) detector.  
A limit of  $G_{\rm M\overline{M}} \le 0.16\,G_{\rm F}$ (90\% C.L.) could be established
\cite{matt91}.
\\
The latest experiment 
at PSI
(Fig. \ref{mmbarsetup}) \cite{prop89,mmbar96}, implemented major improvements 
over the LAMPF setup.
The solid angle for the detection of
the energetic electron was increased by three orders of magnitude 
to 70 \% of $4\pi$
by using a cylindrical magnetic spectrometer equipped with five
concentric proportional chambers and a 64-fold segmented hodoscope which
was constructed from the former SINDRUM~I detector.
The atomic positron is electrostatically accelerated and guided in a momentum 
selective transport system parallel to the   
magnetic field lines to a position sensitive MCP
with resistive anode readout \cite{MCP96}. 
The tracks of these particles 
can be traced back to the interaction region for 
reconstructing a decay vertex providing an additional suppression of 
background. Further the annihilation radiation of the positrons can be observed
in a 12-fold segmented undoped, highly pure CsI crystal calorimeter 
(Fig. \ref{mmbarsetup}).

\begin{figure}[bt] 
  \unitlength 1.0 cm  
  \begin{picture}(11.0,9.0)
   \centering { 
   \hspace{0.9 cm} 
   \mbox{ 
    \psfig{figure=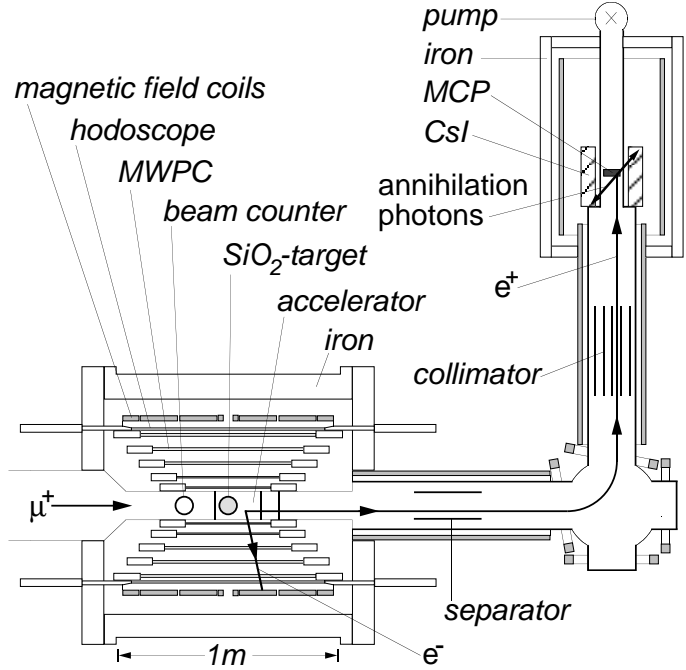,height=9.0cm,angle=90} 
        }
             } 
 \end{picture}
   \label{mmbarsetup}
   \caption[]{
         Top view of the apparatus
        at PSI. 
        The observation of the energetic
        electron from the $\mu ^-$ decay
        in the antiatom in a magnetic spectrometer with a magnetic field
        strength $0.1$~T
        is required in coincidence with
        the detection of the positron,
        which is left behind
        from the atomic shell of the antiatom, on a MCP and at least one
        annihilation
        photon in a CsI calorimeter.
        }
\end{figure}

One of the  design goals for the setup 
was to achieve as high as possible symmetry for the detection of both, 
antimuonium and muonium, in order to reduce the systematic uncertainties 
arising from corrections for efficiencies and acceptances of the 
detector subsystems. The production of the atoms has been
monitored by reversing all electric and magnetic fields regularly 
every few hours for half an hour.

The setup at PSI has a significantly higher sensitivity  
for observing the decay of muonium atoms in vacuum
than to those decaying inside of the production target, 
as it allows the coincident detection of a fast and a slow
particle after the decay. 
Therefore a new determination method for
the muonium production yield could be uniquely exploited. 
It is based on a model established in independent dedicated experiments 
\cite{wood88}, which assumes that the
atoms are produced inside of the SiO$_2$ powder at positions given by the 
stopping distribution of the muons. A one dimensional diffusion process 
describes the escape of the muonium atoms 
into vacuum where their 
velocities follow a 
Maxwell-Boltzmann distribution.
The distribution of time intervals between a stop of a muon
and the detected decay of a muonium atom in vacuum
includes full information on the atom's production rate
(Fig. \ref{Mprod}).
Using an effective
diffusion equation
for the movement of atoms inside the target
parallel to an axis  ($y$) orthogonal to the target
surface the time distribution is
derived for a diffusion length $l=\sqrt{D_M/\lambda_{\mu}}$
which is small compared to the target thickness $a$
\begin{eqnarray}
\nonumber
n_{\rm M{\it vac}}(t) &=&
\frac{f_{\rm M} \cdot exp(-\lambda_{\mu} \cdot t)}
{2 \cdot \sqrt{\pi \cdot D_{\rm M}}} \\
& &\cdot \; \int_0^t dt' \int_0^a dy \:  \frac{S(y) \, (a-y)}{\sqrt{t'^3}} \cdot
exp \left( -\frac{(a-y)^2}{4\cdot D_{\rm M} \cdot t'} \right) \;\;\;,
\end{eqnarray}
with  $D_{\rm M}$ the diffusion constant,
$f_M$ the fraction of muons stopped with a density $S(y)$ inside  
of the target and forming
muonium. There is no muonium in vacuum at $t=0$. 
The maximum of ${n_{\rm M{\it vac}}(t)}$ is approximately at 1.5\,$\mu \rm s$
which is about the average
time of diffusion for the muonium atoms in the target.
The spectra contain a small exponentially decaying 
background arising
from muon decays within the target which can be associated with  the release 
of a secondary electron from the target material.
Numerical fits (Fig. \ref{Mprod})
typically yield a few percent of muonium atoms in vacuum 
with respect to the incoming muons. 
For a flat stopping distribution $S(y)$ and target thickness $a$
large compared to the diffusion length $l$  an analytical integration
results in
\begin{equation}
 n_{\rm M{\it vac}}(t) = f_{\rm M} \;\; exp\left( -\lambda_\mu t\right)
\sqrt{\frac{ D_{\rm M}}{\pi t}} \enspace.
\end{equation}
The expression contains the
diffusion constant $D_{\rm M}$ only as a part of the normalization factor.
\\

\begin{figure}[bt] 
 \unitlength 1.0 cm 
 \begin{picture}(11.8,5.4)
  \centering { 
   \hspace{0.6 cm} 
   \raisebox{-0.6 cm}{
   \mbox{ 
    \psfig{figure=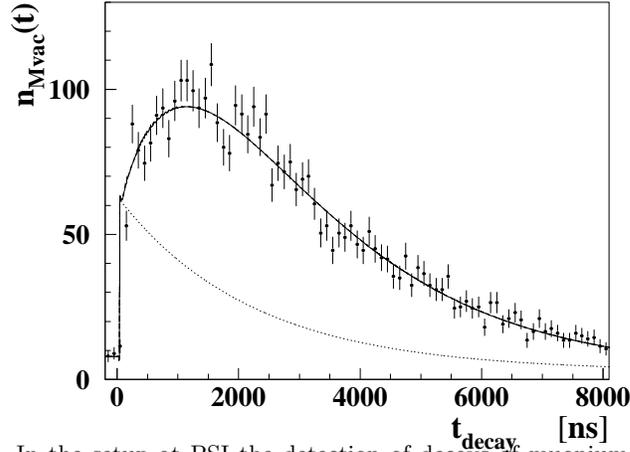,height=6.0cm} 
        }
        }
    }
 \end{picture}                 
 \label{Mprod}
 \caption[]{
        In the setup at PSI the detection of decays of
        muonium atoms in vacuum is favored compared to decays of muons
        inside of the target.
        The distribution of time intervals between the detection of an incoming 
        muon in the beam counter and the observation of the decay of an atom
        in vacuum carries information on the 
        efficiency of the muonium production. 
        The dashed line represents an exponentially decaying background.
        }
\end{figure}

An intermediate result from the PSI experiment, which is carried 
out in a two step approach, is available \cite{mmbar96}
on the basis of an
effective measurement time of 210 hours
during which  $1.4(1)\cdot10^{9}$  muonium atoms decayed inside of  
the fiducial volume
of diameter 9 cm and length 10 cm.
No decay of an antimuonium atom was observed.
There are no entries
in a 20 ns wide window around the expected time of flight of 70 ns
for the positrons from the atomic shell (Fig. \ref{mbarresult}).
The apparent structure around $t_{\rm TOF}-t_{\rm expected} = - 50$ ns
arises from the allowed rare
decay mode $\mu^+ \rightarrow e^+e^+e^- \nu_e \overline{\nu_{\mu}}$ 
in which one
of the positrons is released with low kinetic energy, while the electron is
detected in the magnetic spectrometer.
Due to their significantly higher
initial momenta positrons from these processes arrive
at earlier times at the MCP and can be significantly
distinguished from possible antimuonium decays.
A small part of this observed background signal is due to  
positrons from normal muon decay which experience 
Bhabha scattering in any structural
component in the target region.
%
%
The probability for a conversion
in a 0.1 T magnetic field is 
${P_{\rm M\overline{M}} (0.1~ \rm T) \leq 2.8\cdot 10^{-9}}$
($90\%$ C.L.),
where corrections have been applied to account for 
differences in detection
efficiencies while measuring the muonium
production yield and while searching for antimuonium.

For ${\rm (V\pm A)\times(V\pm A)}$ type interactions,
the conversion probability is suppressed in a magnetic field of 0.1 T
to 35\% of the zero field value. This leads 
through Eq.(\ref{pmm_to_gmm})
to an upper limit of 
$G_{\rm M\overline{M}} \leq  1.8 \cdot 10^{-2}~G _{\rm F}$
(90\% C.L.).

For dileptonic gauge bosons $X^{++}$ in GUT models 
a tight new mass limit of ${ M_{X^{++}}/g_{3l}} > 1.1$~TeV/c$^2$ (90\% C.L.)
can be extracted. This bound exceeds significantly 
the one deducted from high energy Bhabha scattering \cite{petra}.   

With these results from the first step of 
the PSI experiment models with dilepton exchange \cite{fuji94} as 
well as models with heavy
leptons and radiative generation of lepton masses
appear to be less attractive \cite{wong94}. 
This is a nice examples for contributions of research using 
atomic objects for
solving problems in the domain of particle physics.

\begin{figure}[bt]

 \unitlength 1.0 cm 
 \begin{picture}(11.0,4)
 \raisebox{-0.3cm}{
 \centering { 
   \hspace{-0.5 cm} 
   \mbox{ 
    \psfig{figure=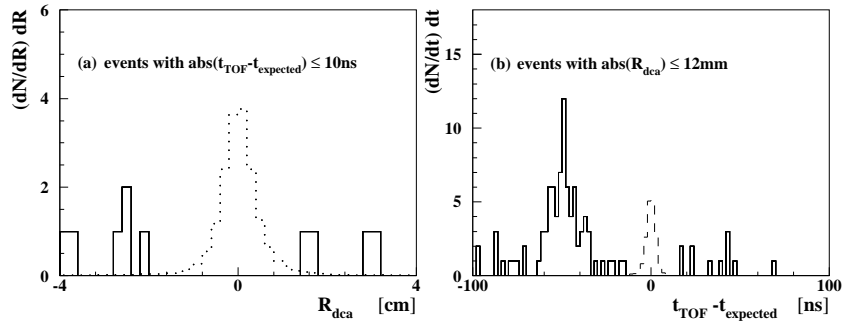,width=11.0cm} 
        }
    }
 } 
 \end{picture}
   \centering\caption[]
        {\label{mbarresult}
         The number of events with identified energetic electron
         and slow positron as a function of
         (a) the distance of closest approach $R_{\rm dca}$
         between the electron track in the
         magnetic spectrometer and the back projection of the position
         measured at the MCP and (b) the difference of the
         positron's time of flight $t_{\rm TOF}$
         to the expected arrival time $t_{\rm expected}$.
         The signal at earlier times is due
         to the allowed decay channel
         $\mu \rightarrow  3e2\nu$ and Bhabha scattering.
         It is smeared out because of the different
         acceleration voltages used.
         No event satisfied the
         required coincidence signature.
         The dotted and dashed curves correspond to a simulated
         signal for ${ G_{\rm M\overline{M}}}=0.05\,G_{\rm F}$.
         }
\end{figure}

\section{Outlook for Future Experiments}

The measurements in the second stage of 
the experiment at PSI promise further advances.
Among the major improvements are a detector for positrons  
with four times enhanced efficiency \cite{MCP96}
and a beam line ($\pi$E5) with 5 times
higher muon flux.
Data have been collected for some 1300  hours and
a preliminary result \cite{Schmidt97} is available which
sets in ${\rm (V\pm A)\times(V\pm A)}$ coupling an upper limit of 
${\rm G_{M\overline{M}}} \leq  3.2 \cdot 10^{-3}~\rm G _{\rm F}$
(90\%C.L.).
This
provides an even more stringent test
for speculative extensions to the standard model,
in particular to the left-right symmetric models predicting a lower bound
on $\rm G_{M\overline{M}}$ \cite{herc92}.

In order to increase the sensitivity 
of detecting
a possible conversion process of muonium into antimuonium
considerably in the future a new experimental approach will be required.
The present setup has come close to the limits imposed by the available
muon fluxes at present meson factories, realistic durations of running times
and the rate capabilities of available proportional wire chambers.
The number of accidental coincidences is expected to 
become a serious problem for higher beam rates.
At present, possible improvements to the existing setup at PSI 
appear to promise only marginal progress.
However, at future highly intense pulsed muon sources
one could take advantage of the time evolution of the conversion 
signal which increases quadratically in time (Eq. {\ref{t_oscill}).
Therefore a detection scheme could be envisaged which again uses the powerful
coincidence detection of both constituents of antimuonium 
and which starts to look for antimuonium decays  a few muon lifetimes after
the formation of the system.

It should be noted that final state interaction in muonium 
decay could mimic an antimuonium decay, when energy is transfered 
from the positron of the $\mu^+$ decay to the electron in the atomic
shell (internal Bhabha scattering). An energy transfer of more than 
10\,MeV while the positron remains with less than 0.1\,MeV kinetic energy
has a probability of well below 10$^{-11}$ \cite{prop89}.
However, this process can be distinguished from an antimuonium decay by the 
characteristic energy spectra of the detectable 
particles as in the case of potential
background from the allowed $\mu \rightarrow 3e2\nu$ decay. 
Therefore, there is no principle limitation which could prevent
much more sensitive searches beyond the present bounds.

\section*{Acknowledgements}
We are indebted to Prof. G. zu Putlitz for his constant support, advice
and encouragement during our work on this appealing subject in the
framework of international collaboration. \\ 

\vspace*{1mm}
\noindent{
$^*$ {\footnotesize { Present address: 
Department of Physics, Massachusetts Institute of Technology, 
\hspace*{3mm} Cambridge, MA. 02139} }}

\bibliographystyle{plain}

\end{document}